# Hexagonal boron nitride nanophotonics

Grudinin D.V.[1*], Ermolaev G.A.[1*], Baranov D.G.[1], Toksumakov A.N.[1], Voronin K.V.[2], Slavich A.S.[1], Vyshnevyy A.A.[1], Mazitov A.B.[3], Kruglov I.A.[1], Ghazaryan D.[4], Arsenin A.V.[1,4], Novoselov K.S.[5,6,7], and Volkov V.S.[1]

[1]*Emerging Technologies Research Center, XPANCEO, Dubai Investment Park First, Dubai, United Arab Emirates*

[2]*Donostia International Physics Center (DIPC), 20018 Donostia-San Sebastián, Spain*

[3]*Institute of Materials, École Polytechnique Fédérale de Lausanne, 1015 Lausanne, Switzerland*

[4]*Laboratory of Advanced Functional Materials, Yerevan State University, Yerevan 0025, Armenia*

[5]*National Graphene Institute (NGI), University of Manchester, Manchester, M13 9PL, UK*

[6]*Department of Materials Science and Engineering, National University of Singapore, Singapore, 03-09 EA, Singapore*

[7]*Chongqing 2D Materials Institute, 400714, Chongqing, China*

*These authors contributed equally

e-mail: vsv@xpanceo.com

## ABSTRACT

**A global trend to miniaturization and multiwavelength performance of nanophotonic devices drives research on novel phenomena, such as bound states in the continuum and Mietronics, as well as the survey for high-refractive index and strongly anisotropic materials and metasurfaces. Hexagonal boron nitride (hBN) is one of promising materials for the future nanophotonics owing to its inherent anisotropy and prospects of high-quality monocrystals growth with atomically flat surface. Here, we present highly accurate optical constants of hBN in the broad wavelength range of 250–1700 nm combining the imaging ellipsometry measurements scanning near-field optical microscopy and first-principle quantum mechanical computations. hBN's high refractive index, up to 2.75 in ultraviolet (UV) and visible range, broadband birefringence of ~0.7, and negligible optical losses make it an outstanding material for UV and visible range photonics. Based on our measurement results, we propose and design novel optical elements: handedness-preserving mirrors and subwavelength waveguides with dimensions of 40 nm operating in the visible and UV range, respectively. Remarkably, our results offer unique opportunity to bridge the size-gap between photonics and electronics.**

## INTRODUCTION

The rise of all-dielectric nanostructures[1], consisting of Mie-resonant nanoparticles[2], makes clear the importance of lossless high refractive index materials[3]. For a given wavelength of light $\lambda$, it is the refractive index $n$ that controls the minimum size (~ $\lambda/n$) of nanostructure features, boosts field enhancement factor, and resonance figure of merit[4]. In turn, the absence of loss guarantees the presence of high-quality resonance and/or secures long-distance light propagation[5]. For this purpose, researchers usually leverage silicon (Si)[6], gallium phosphide (GaP)[7], or titanium dioxide ($TiO_2$)[8].

At the same time, recent works[9-12] have revealed that layered materials such as transition metal dichalcogenides (TMDs)[13] could also serve as high refractive index materials. Furthermore, the layered structure, unlike in the case of traditional semiconductors (Si, GaP, and $TiO_2$), essentially produces a giant optical anisotropy[14], which is now recognized an indispensable degree of freedom. As a result, TMDs enable numerous photonic phenomena: extreme skin-depth engineering[14], exciton-polariton transport[15], Zenneck surface waves[16], and sophisticated anapole states[11]. Nevertheless, TMDs strongly absorb light in the visible and ultraviolet spectral intervals, which severely limits their operation range.

By contrast, hexagonal boron nitride (hBN) is a promising wide-bandgap material[17], as it consolidates all three advantages: broadband zero optical losses, high-index, and giant optical anisotropy. Besides, hBN is an essential building block for two-dimensional electronic and optoelectronic devices[18], being widely used as a high-quality dielectric and encapsulation material[19]. Despite the wide use of hBN in nanophotonics and optoelectronics, only a few studies[20-22] have determined its anisotropic optical response and in the relatively narrow wavelength ranges of 245–310 nm[20], 220–780 nm[21], 1530 nm[22]. Furthermore, the optical properties were studied via transmittance spectroscopy, micro-reflectance, which have a limited accuracy, and the near-field microscopy, which yields a result at a single wavelength. Therefore, highly accurate broadband optical properties of hBN measured by ellipsometry are in strong demand. However, the typical flake size of hBN is much lower than required for most ellipsometers, which makes ellipsometry challenging.

In this work, we performed the first broadband (250–1700 nm) ellipsometric measurement of the anisotropic dielectric function of hBN by the imaging ellipsometer that can collect the signal from a single flake[14]. We refined and validated the measured optical properties via a combined experimental and theoretical study which included: micro-transmittance[23], scattering near-field optical microscopy[14,22], and the first-principle quantum mechanical computations[14,24]. We discovered that hBN has no losses and giant birefringence ($\Delta n \sim$ 0.7) throughout the whole spectral range. As a result, hBN complements titanium dioxide in the deep-, mid-, and near-ultraviolet intervals as high-index material. The acquired optical properties allow hBN to take its place among other high refractive index anisotropic materials, including $MoS_2$[14], $BaTiS_3$[25], and α-$MoO_3$[26]. hBN enables the next-generation high-performance photonic elements, operating at visible and ultraviolet frequencies[27], which we demonstrate by designing a handedness-preserving mirror and a 50-nm-wide subwavelength waveguide with an unprecedented integration density.

# RESULTS

## Far- and near-field studies of hBN's dielectric tensor

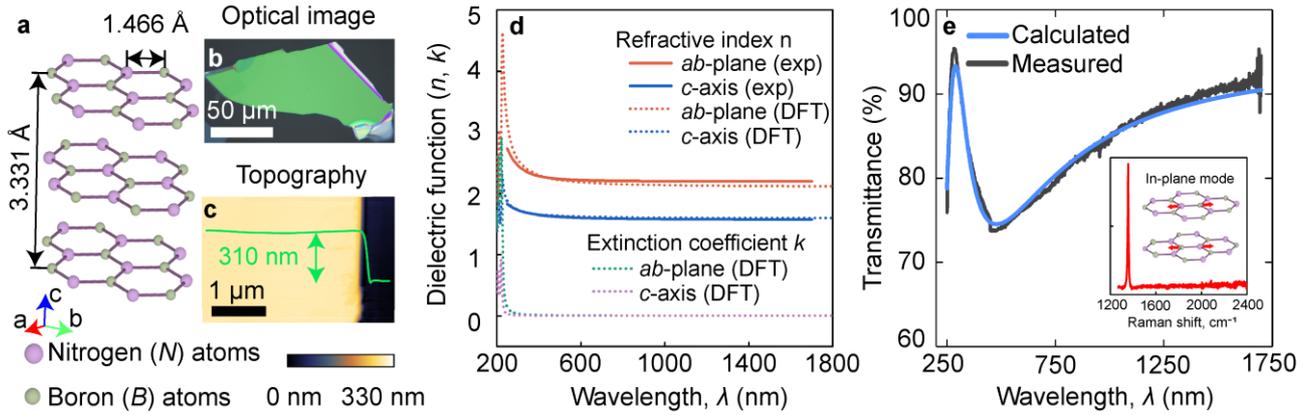

**Figure 1. Optical properties of hBN. a** Crystal structure of hBN. **b** Optical and **c** AFM images of hBN flake. **d** Refractive index ($n$) and extinction coefficient ($k$) along the crystallographic *ab*-plane and *c*-axis. DFT (density functional theory) in labels refers to the first-principle computations. **e** Measured (black line) and calculated (blue line) transmittance spectra. The inset shows Raman spectra taken from the flake presented in (b).

Figure 1a shows the crystal structure of hBN. Similar to many other layered materials (MoS$_2$, WS$_2$, *etc.*), hBN has a honeycomb lattice within the layers bound by the weak van der Waals forces. Therefore, hBN has a diagonal dielectric tensor in the form diag($n_{ab}^2$, $n_{ab}^2$, $n_c^2$), where $n_{ab}$ and $n_c$ are the complex refractive indices along the crystallographic *ab*-plane and *c*-axis, respectively (see Figure 1a)[14].

The measurement of hBN anisotropic dielectric tensor is a challenging task owing to numerous experimental difficulties. First of all, etalon data can be acquired only with exfoliated flakes. Indeed, exfoliation produces monocrystals with a minimum number of defects[28], whereas chemically[9] or physically[29] synthesized thin films have a polycrystalline structure and numerous defects. However, exfoliation, at best, could provide a sample size of only a hundred micrometers (Figure 1b), which severely limits the choice of spectroscopy techniques. Secondly, standard transmittance and reflectance measurements probe only the in-plane dielectric response and are not sensitive to the out-of-plane component[30,31]. Fortunately, the imaging spectroscopic ellipsometer[14] can solve both problems. It is a hybrid of an optical microscope and an ellipsometer, which allows us to record the local optical signal (microscope advantage), and at oblique angles, thereby probing the full dielectric tensor (ellipsometry advantage).

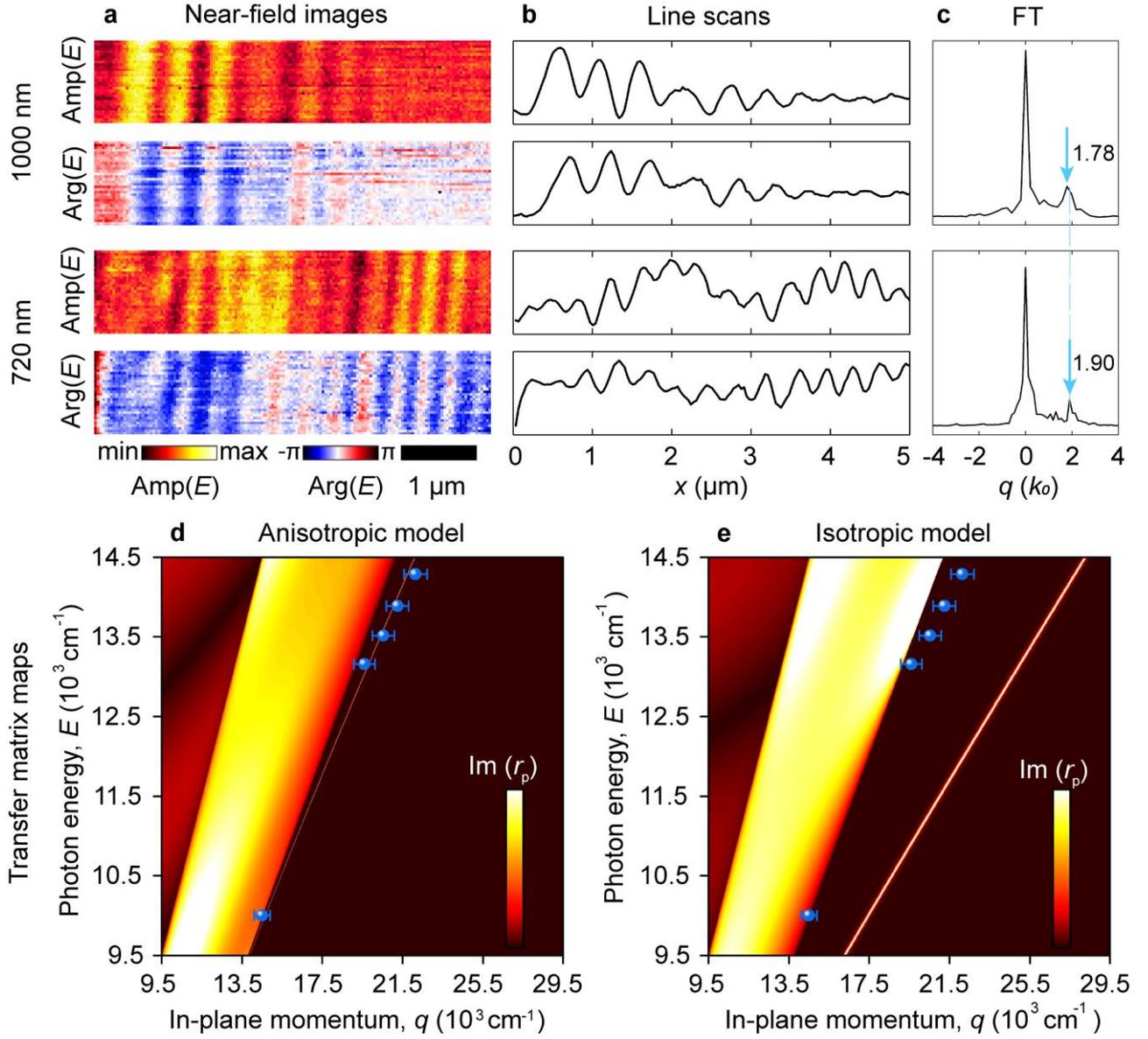

**Figure 2. Near-field analysis of hBN. a** Near-field images, real part Re(*E*) and phase Arg(*E*), of the electric field *E* taken at 1000 nm (top) and 720 nm (bottom). **b** *x*-line scans taken from panel (a) and averaged over 1.1 μm along the *y*-axis (other wavelength images are collected in Supplementary Note 2). **c** Fourier transform (FT) amplitude of the complex near-field signal in (b). Green arrows mark the peak associated with the waveguide mode. **d, e** Transfer matrix calculations for hBN/SiO$_2$ system for anisotropic and isotropic (in the assumption of optical constants for the *c*-axis equal to those for the *ab*-plane) hBN optical models. The experimental ($q = 1/\lambda$, $E = hc/\lambda$) data points (blue circles) show good agreement with the calculated dispersion based on the anisotropic dielectric function presented in Figure 1d.

Supplementary Note 1 shows experimental spectra of ellipsometric parameters *Ψ* (amplitude) and *Δ* (phase). Note that we chose a thick sample of 310 nm to identify the transparency region (see Figure 1c). Remarkably, one can notice pronounced Fabri-Perot oscillations in *Ψ* and *Δ* (Supplementary Note 1) in the whole measured spectral interval (250–1700 nm). It implies zero optical losses of hBN in this range. Therefore, Cauchy model[14,15] describes both parts ($n_{ab}$ and $n_c$) of the dielectric tensor (see Methods). Figure 1d shows the resulting dielectric function in comparison with the first-principle calculations (see Methods). Experimental optical constants have a perfect match with the theoretical values (see Figure 1d). As an additional verification, we also recorded transmittance spectra, which yield a good agreement (see Figure 1e) with the transfer matrix calculations[32] based on dielectric function from Figure 1d, whereas the Raman measurements (see the inset of Figure 1e) confirm the hexagonal structure of hBN[33]. However, far-field measurements can only confirm the in-plane component of the dielectric tensor.

For this reason, we also analyzed the near-field distribution of the signal scattered by hBN flake, which forms a planar waveguide, using a scattering-type near-field optical microscope (s-SNOM). The effective mode index in such configuration strongly depends on both the in-plane and the out-of-plane dielectric functions allowing us to conduct an unambiguous validation. We used incident wavelengths in the range of 700–1000 nm to excite the mode inside the structure by focusing the light on the tip of s-SNOM. The excited cylindrical waves propagate inside the hBN flake and scatter at the edge of the flake. That scattered field interferes with the background signal thus giving the oscillating pattern of the near-field amplitude signal (see Figure 2a-b). To retrieve the effective mode index from the measurements we analyzed the near-field phase and amplitude simultaneously by studying the complex Fourier transform (FT) (see Figure 2c). The FT has two pronounced peaks: the one around $q$=0 corresponds to the strong tip-sample interaction, meanwhile, the one at finite $q$ corresponds to the waveguide mode. Note that the observed effective mode index is different from the real effective index of the mode, which is a consequence of the specific geometry of the experiment[14,15,22]. The shift from the actual mode is described by the following relation:

$$n_{\text{eff}} = n_{\text{SNOM}} - \cos(\alpha)\sin(\beta), \qquad (1)$$

where $n_{\text{eff}}$ is the effective index of the mode, propagating inside the planar waveguide, $n_{\text{SNOM}}$ is the effective index obtained from the near-field measurements, $\alpha$ = 45° is the angle between the illumination wavevector and its projection on the sample's surface plane and $\beta$ = 30° is the angle between the projection of the wavevector and the flake edge. Additionally, we did not observe standing waves (reflection from the edge) during the near-field measurements. Otherwise, the observable effective mode index would be doubled compared to the effective index of the propagating mode and would be around 3[34], which is clearly not the case, as seen from Fourier spectra in Figure 2c.

Given the extracted refractive index components $n_{ab}$ and $n_c$, we calculated the energy ($E = hc/\lambda$)–momentum ($q = 1/\lambda$) dispersion relation of the waveguide mode using the transfer matrix method[32]. The experimental data points (blue circles) can be observed on top of the dispersion color map (see Figure 2d-e). Apparently, experimental points agree perfectly with the anisotropic model (see Figure 2d) and sharply disagree with an isotropic model (see Figure 2e) where $n_c$ is set to be equal to $n_{ab}$. Consequently, the excellent agreement between the experimental and theoretical dispersion unambiguously validates our dielectric permittivity of hBN, presented in Figure 1d.

**Comparison of hBN's optical response with other materials**

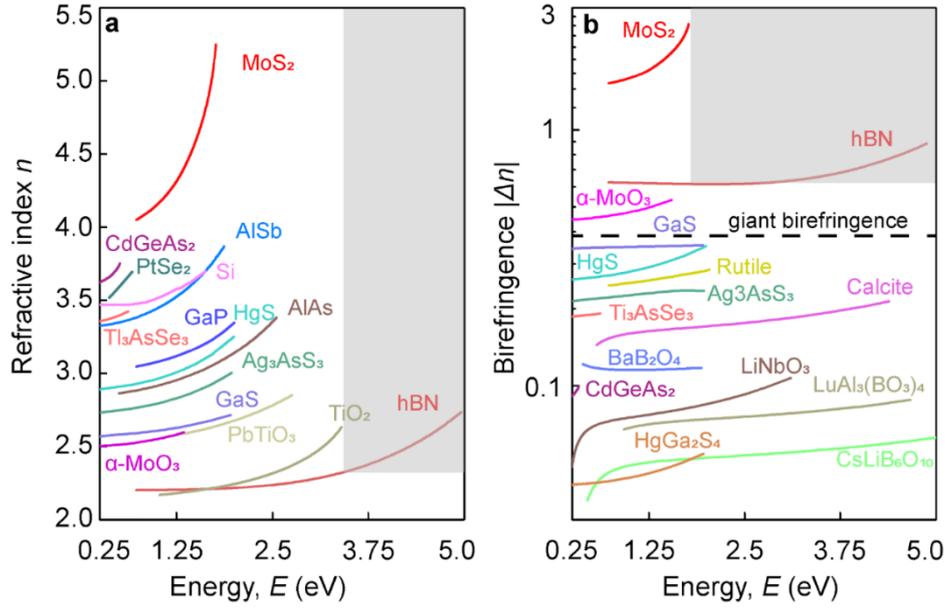

**Figure 3. hBN in the core of high-refractive-index birefringent materials. a** Refractive index and **b** birefringence of comparison of conventional materials in their respective transparency windows. Gray areas in the panel's spectral ranges, where hBN outperforms other materials in refractive index and birefringence, making it the best candidate for photonic applications.

It is worth comparing hBN optical properties with other materials (see Figure 3). Firstly, hBN transparency region is the widest among traditionally used and novel materials. As a result, it fills the spectral gap for short wavelengths (gray areas in Figure 3) and favors hBN in broadband all-dielectric nanophotonics. Secondly, its refractive index is higher than 2 in the visible and infrared spectral intervals and approaches 2.75 at the ultraviolet wavelengths. Thus, hBN enriches the list of high-index materials. Finally, hBN birefringence outperforms most anisotropic materials and thereby complements the library of materials with giant optical anisotropy. Altogether these properties create a firm foothold for advanced photonic engineering, as demonstrated in the next section, especially for ultraviolet and visible frequencies.

**Handedness-preserving mirrors and subwavelength waveguides**

Reasonably high refractive index (~ 2.2) and negligible absorption loss in the visible range make hBN an interesting material for a variety of optical applications. As an example of a potential application, we demonstrate the performance of a handedness-preserving mirror. By this term we refer to a planar structure that, at a particular wavelength, can reflect a circularly polarized plane wave of a given handedness without flipping its handedness but is transparent for the wave of opposite handedness at the same wavelength[35,36].

To that end, we utilize the design proposed by Semnani and coworkers[37]. The structure is presented by a dielectric photonic crystal slab with a square lattice and its unit cell containing a circular hole in the middle and two satellite elliptical holes, shown in Figure 4a. Note that we modified the structure suggested in[37] by introducing the angle $\vartheta$ of the elliptical holes (see Figure 4a). It gives an additional degree of freedom to achieve optimal polarization responses. As a result of this arrangement (Figure 4a), the structure possesses only the mirror symmetry with respect to the horizontal plane, $M_{xy}$, and two-fold rotational symmetry around the $z$-axis $C_z^2$. These symmetries are crucial for the desired polarization response of the structure[38].

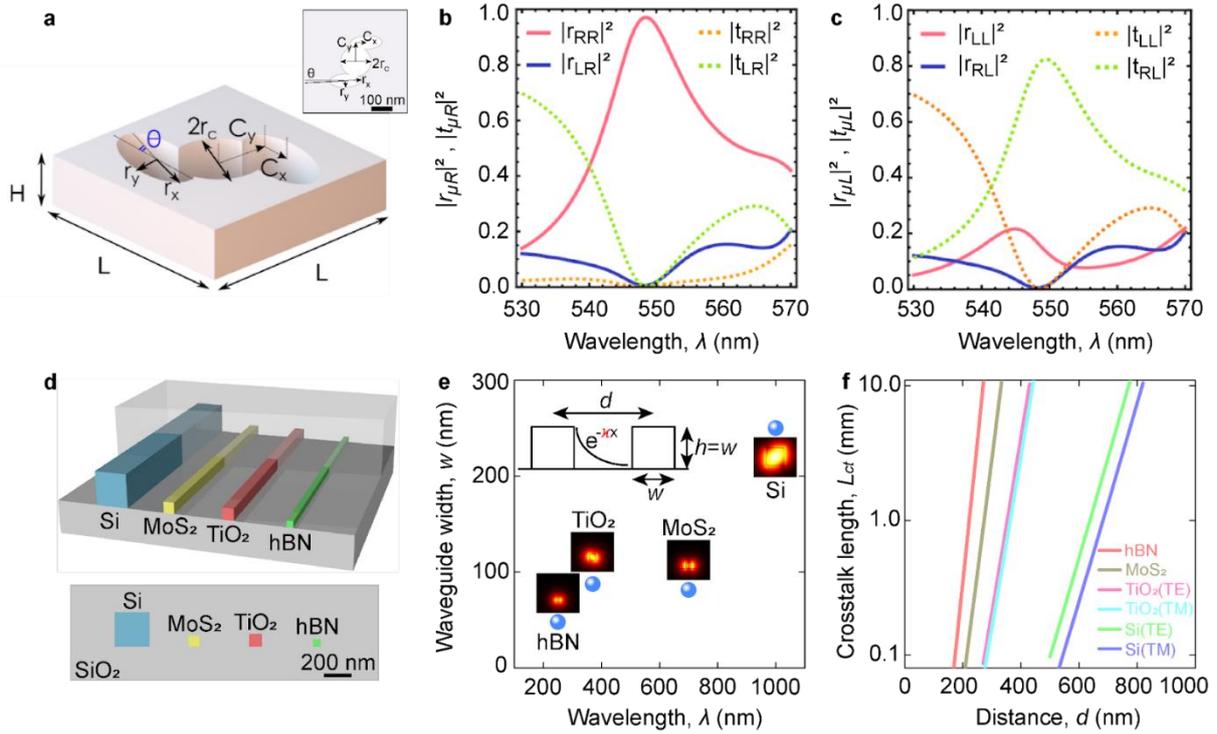

**Figure 4. Prospective applications of high refractive index birefringent hBN.** Handedness-preserving mirror **a-c** and subwavelength waveguide **d-e**. **a** Geometry of the unit cell of the dielectric handedness-preserving mirror. **b** Spectra of intensity reflection and transmission coefficients of the optimal handedness-preserving mirror made of a hBN film in the circular polarization basis for a right-handed incident plane wave. **c** The same as (b) for a left-handed incident plane wave. **d** Schematic comparison of the dimensions of waveguides made from high-refractive index materials: 48×48 nm² for hBN, 81×81 nm² for MoS$_2$, 88x88 nm² for TiO$_2$, and 250×250 nm² for Si. **e** Waveguides width for different materials with constant $\varkappa$ = 2.25 μm$^{-1}$. **f** Relation of crosstalk length and distance between cores of the waveguides for high-refractive index materials. The width is optimized for each point.

We illustrate the applicability of hBN for this purpose by designing such a handedness-preserving mirror at a target wavelength of 550 nm. Numerical optimization was performed with the use of particle swarm algorithm in a finite-difference time domain (FDTD) commercial solver (Lumerical). By enforcing sub-diffraction periodicity of the mirror, we obtain the set of unit cell dimensions yielding near-perfect polarization conversion for normally incident light at the target wavelength: $L$ = 490 nm, $H$ = 224 nm, $r_c$ = 70 nm, $r_x$ = 147 nm, $r_y$ = 64 nm, $C_x$ = 49 nm, $C_y$ = 91 nm, $\vartheta$ = −3°. With this set of dimensions, the mirror reflects normally incident right-handed (RH) plane wave back into a RH wave with near-unity efficiency, $|r_{RR}|^2 \approx 0.96$ (see Figure 4b). At the same time, a left-handed (LH) incident wave is nearly perfectly transmitted into a RH wave, $|t_{RL}|^2 > 0.8$, as seen from Figure 4c.

Aside from chiral optics, we envision the usage of hBN in integrated nanophotonics[38], a promising technology for efficient parallel data processing required for artificial intelligence systems[40]. However, integrated photonic elements are still much larger compared to electronic counterparts, due to the diffraction limit imposed by the wave nature of light. Hence, to make photonic elements compact, one should operate close to the minimum wavelength of the material transparency window, which is the smallest for hBN (see Figure 2).

To assess hBN as photonic material, we studied the performance of the photonic waveguides based on hBN and compared them with the waveguides based on other high-index materials, such as MoS$_2$, TiO$_2$, and Si (see Figure 4d). We considered waveguides operating at the lowest possible wavelength $\lambda$ (250, 370, 700, and 1000 nm for hBN, TiO$_2$, MoS$_2$, and Si, respectively), where the core material optical losses are negligible. For material comparison, we limited our analysis to square-shape waveguides with a given exponential decay $\varkappa$ = 2.25 μm$^{-1}$ (see inset in Figure 4e) of mode field in the cladding (SiO$_2$), defined as:

$$\varkappa = \frac{\sqrt{n_{\text{eff}}^2 - n_s^2}}{\lambda}, \quad (2)$$

where $n_{\text{eff}}$ the effective index of the mode, $n_s$ is refractive index of the waveguide cladding. Such a strong decay ensures relatively low crosstalk noise. Under these constraints, we concluded that hBN waveguide width can be as small as 47 nm, bringing the size scale of photonic elements down to that of electronic components (see Figure 4e). This surprising result originates from the low operation wavelength ($\lambda \sim 250$ nm), high refractive index ($n \sim 2.8$), and giant anisotropy ($\Delta n \sim 0.9$). Notably, $MoS_2$ outperformed $TiO_2$ (see Figure 4e) despite its twice as high operating wavelength thanks to its much higher refractive index and stronger optical anisotropy.

The integration density of photonic elements is not limited by their physical dimensions but rather by the evanescent field coupling between adjacent waveguides. When the lateral distance between waveguides $d$ decreases, the crosstalk distance $L_{ct}$ characterizing the coupling strength decreases exponentially (Figure 4f). During calculations, the width of the square-shaped waveguide was optimized to maximize $L_{ct}$. Again, hBN and $MoS_2$ demonstrate the best results, owing to their high refractive index and giant anisotropy. For instance, hBN waveguides with a 10 mm crosstalk length can only be separated by 270 nm. Hence, a 1-mm chip can accommodate up to 3700 separate waveguides, operating at a high frequency ($\nu = c/\lambda = 1.2 \times 10^{15}$ Hz). As a result, photonic integrated circuits based on such waveguides can potentially revolutionize integrated photonics and semiconductor industry.

## DISCUSSION AND CONCLUSION

High refractive index, optical anisotropy, and transparency are the main figures of merit for the materials which define modern nanophotonics. Novel natural materials combining all three properties are in high demand. In this work, we present hBN as a promising material for the visible and ultraviolent range. We determined the exact values of anisotropic dielectric permittivity tensor of hBN in the broad spectral range (250–1700 nm) using cross-validation of far- and near-field techniques, accompanied by first-principle calculations. Our results show high refractive index, transparency over the whole studied spectral range and giant optical anisotropy of $\Delta n \sim 0.7$. Furthermore, we suggest applications of hBN for chiral optics and integrated photonics. We propose designs of handedness-preserving mirrors and subwavelength waveguides based on hBN. In the case of waveguides, hBN shows outstanding performance: the waveguide width can be as small as 40 nm with the integration density exceeding 3000 elements per mm. Overall, our results demonstrate that hBN can become a key material for future ultraviolet nanophotonics.

## AUTHOR CONTRIBUTIONS

G.D.V. and E.G.A. contributed equally. V.V.S., N.K.S., A.A.V., G.D., B.D.G., and E.G.A. suggested and directed the project. T.A.N. and G.D. prepared the samples. G.D.V., S.A., T.A.N., and E.G.A. performed the measurements. G.D.V., E.G.A., B.D.G., V.A.A., M.A.B., K.I.A., and V.K.V. provided theoretical support. G.D.V., E.G.A., and D.G.B. wrote the original manuscript. G.D.V., E.G.A., B.D.G., V.A.A., G.D., A.A.V., N.K.S., and V.V.S. reviewed and edited the paper. All authors contributed to the discussions and commented on the paper.

## COMPETING INTERESTS

The authors declare no competing interests.

# METHODS

**Sample preparation.** Bulk hBN crystals were purchased from 2D Semiconductors (Scottsdale, USA) and exfoliated on top of required substrates: Si/SiO$_2$ and CaF$_2$. Prior to exfoliation, the corresponding substrates were subsequently cleaned in acetone, isopropanol alcohol, and deionized water. Then, those were subjected to oxygen plasma (O$_2$) to remove the ambient adsorbates. Following the plasma cleaning, substrates were put on a hot plate at the temperatures of 120–140 °C for 2 minutes. After the heating, the scotch-tape from Nitto Denko Corporation (Osaka, Japan) with loaded crustal hBN was brought in contact with substrates. Afterwards, the tape was removed completing the exfoliation procedure.

**Imaging ellipsometry.** For determination of hBN anisotropic dielectric tensor, we used Accurion nanofilm_ep4 ellipsometer. Measurements were performed in a broad spectral range 250–1700 nm in steps of 1 nm at several incident angles 45°, 50°, and 55° with 7× objective. For optical constants dispersion analysis, we implemented Cauchy model $n(\lambda) = A+B/\lambda^2+C/\lambda^4$, where *A*, *B*, and *C* are Cauchy fitting parameters, for both the in-plane and the out-of-plane components.

**First-principle calculations.** First-principles calculations of optical constants of hBN were performed within the Vienna Ab Initio Simulation Package (VASP)[41,42]. First, the atomic positions of the crystal were relaxed using standard density functional theory (DFT) approach until the interatomic forces were less than $10^{-3}$ eV/Å. Second, the initial Kohn-Sham orbitals were calculated with electronic self-consistent cycle convergence criteria of $10^{-8}$ eV. Finally, the real and imaginary parts of frequency-dependent dielectric function were obtained using single-shot GW calculation[43]. Cutoff energy of the plane waves basis set was set to 500 eV, and the *Γ*-centered 15×15×5 k-points mesh was used to sample the first Brillouin zone. The GW-type projector-augmented wave pseudopotentials[44] were used to describe the behavior of wavefunctions in the core region and the exchange-correlation effects of electrons.

**Raman spectroscopy.** The Raman spectra were measured with a confocal scanning Raman microscope Horiba LabRAM HR Evolution (HORIBA Ltd., Kyoto, Japan) equipped with a 632.8 nm linearly polarized laser, 600 lines/mm diffraction grating, and ×100 objective (N.A. = 0.90) in a spectra range of 1200–2400 cm$^{-1}$. The spectra were recorded with 0.7 mW incident laser power, spot size of ~ 0.86 μm and integration time of 10 s with 10 spectra accumulation. Unpolarized detection was used to provide a higher signal-to-noise ratio.

**Microtransmittance spectra.** Accurion nanofilm_ep4 ellipsometer was used for microtransmittance measurements at normal incidence for 250–1700 nm in steps of 1 nm. Input and output ellipsometer branches were set vertically with polarizer, compensator, and analyzer positions at 0°. During the measurements at each wavelength, we recorded the intensity image and used only the part with the sample. For reference we also measured air transmittance intensity.

**Near-field optical nano-spectroscopy.** Near-field imaging was performed with the use of commercially available scattering scanning near-field optical microscope (s-SNOM, neaSNOM www.neaspec.com). For sample illumination we used a monochromatic tunable Ti:Sapphire laser (www.avesta.ru) of the wavelength from 700–1000 nm spectral interval. The microscope was used in the reflection mode, which means that to focus the light on and collect the signal from the probe-sample interface was used the same parabolic mirror (Supplementary Note 2). The near-field signal was filtered from the raw detector data using a pseudo-heterodyne interferometer at high harmonics (the third one). For the measurement was used a Platinum/Iridium5 (PtIr$_5$) coated AFM tip (ARROW-NCPt-50, https://www.nanoworld.com) with resonant frequency of around 275 kHz and tapping amplitude 47 nm.

**Numerical calculations for handedness-preserving mirrors.** Numerical simulations of the handedness-preserving mirror were performed using commercial finite-difference time-domain solver (Lumerical). The structure was excited with two normally-incident plane wave sources emulating either a right-handed, or a left-handed circularly polarized plane wave. Elements of the scattering matrix in the basis of circular polarizations were obtained by calculating the overlap of the scattered (reflected and transmitted) field with the fields of right-handed and left-handed waves.

**Numerical calculations for waveguides.** Numerical calculations for waveguides width with constant $\varkappa$ and waveguides crosstalk length were performed using a combination of COMSOL Multiphysics and MATLAB via live link. COMSOL was used as a solver, for every task a simple 2D geometry model was used. MATLAB was used to optimize the geometry for the solver, finding exact width of the waveguide for given $\varkappa$ and optimizing the width of the waveguides to achieve maximum crosstalk length with given distance between centers of the cores. For the first task a simple binary search algorithm was used. For the second task we used modification of a gradient descent algorithm.

## DATA AVAILABILITY

The datasets generated during and/or analyzed during the current study are available from the corresponding author on reasonable request.

## REFERENCES

1. Kivshar, Y. The Rise of Mie-tronics. *Nano Lett.* **9**, 3513–3515 (2022).

2. Evlyukhin, A. B. *et al.* Demonstration of magnetic dipole resonances of dielectric nanospheres in the visible region. *Nano Lett.* **12**, 3749–3755 (2012).

3. Khurgin, J. B. Expanding the Photonic Palette: Exploring High Index Materials. *ACS Photonics* vol. **9** 743–751 (2022).

4. Baranov, D. G. *et al.* All-dielectric nanophotonics: the quest for better materials and fabrication techniques. *Optica, OPTICA* **4**, 814–825 (2017).

5. Atabaki, A. H. *et al.* Integrating photonics with silicon nanoelectronics for the next generation of systems on a chip. *Nature* **556**, 349–354 (2018).

6. Evlyukhin, A. B., Matiushechkina, M., Zenin, V. A., Heurs, M. & Chichkov, B. N. Lightweight metasurface mirror of silicon nanospheres. *Optical Materials Express* **10**, 2706 (2020).

7. Khmelevskaia, D. *et al.* Directly grown crystalline gallium phosphide on sapphire for nonlinear all-dielectric nanophotonics. *Applied Physics Letters* vol. 118 201101 (2021).

8. Ermolaev, G. A., Kushnir, S. E., Sapoletova, N. A. & Napolskii, K. S. Titania Photonic Crystals with Precise Photonic Band Gap Position via Anodizing with Voltage versus Optical Path Length Modulation. *Nanomaterials* **9**, 651 (2019).

9. Ermolaev, G. A. *et al.* Broadband optical properties of monolayer and bulk $MoS_2$. *npj 2D Materials and Applications* **4**, 21 (2020).

10. Green, T. D. *et al.* Optical material anisotropy in high-index transition metal dichalcogenide Mie nanoresonators. *Optica* **7**, 680 (2020).

11. Verre, R. *et al.* Transition metal dichalcogenide nanodisks as high-index dielectric Mie nanoresonators. *Nat. Nanotechnol.* **14**, 679–683 (2019).

12. Popkova, A. A. *et al.* Nonlinear Exciton-Mie Coupling in Transition Metal Dichalcogenide

Nanoresonators. *Laser & Photonics Reviews* **16**, 2100604 (2022).

13. Ermolaev, G. *et al.* Topological phase singularities in atomically thin high-refractive-index materials. *Nat. Commun.* **13**, 2049 (2022).

14. Ermolaev, G. A. *et al.* Giant optical anisotropy in transition metal dichalcogenides for next-generation photonics. *Nat. Commun.* **12**, 854 (2021).

15. Hu, F. *et al.* Imaging exciton–polariton transport in $MoSe_2$ waveguides. *Nature Photonics* **11**, 356–360 (2017).

16. Babicheva, V. E. *et al.* Near-Field Surface Waves in Few-Layer $MoS_2$. *ACS Photonics* **5**, 2106–2112 (2018).

17. Caldwell, J. D. *et al.* Photonics with hexagonal boron nitride. *Nature Reviews Materials* **4**, 552–567 (2019).

18. Dean, C. R. *et al.* Boron nitride substrates for high-quality graphene electronics. *Nat. Nanotechnol.* **5**, 722–726 (2010).

19. Bandurin, D. A. *et al.* Cyclotron resonance overtones and near-field magnetoabsorption via terahertz Bernstein modes in graphene. *Nature Physics* vol. **18** 462–467 (2022).

20. McKay, M. A., Li, J., Lin, J. Y. & Jiang, H. X. Anisotropic index of refraction and structural properties of hexagonal boron nitride epilayers probed by spectroscopic ellipsometry. *Journal of Applied Physics* **127**, 053103 (2020).

21. Segura, A. *et al.* Natural optical anisotropy of h-BN: Highest giant birefringence in a bulk crystal through the mid-infrared to ultraviolet range. *Physical Review Materials* vol. **2**, 024001 (2018).

22. Hu, D. *et al.* Probing optical anisotropy of nanometer-thin van der waals microcrystals by near-field imaging. *Nature Communications* **8**, 1471 (2017).

23. Frisenda, R. *et al.* Micro-reflectance and transmittance spectroscopy: a versatile and powerful tool to characterize 2D materials. *Journal of Physics D: Applied Physics* **50**, 074002 (2017).

24. Álvarez-Pérez, G. *et al.* Van der Waals Semiconductors: Infrared Permittivity of the Biaxial van der Waals Semiconductor α-$MoO_3$ from Near- and Far-Field Correlative Studies. *Advanced Materials* **32**, 2070220 (2020).

25. Niu, S. *et al.* Giant optical anisotropy in a quasi-one-dimensional crystal. *Nature Photonics* **12**, 392–396 (2018).

26. Ma, W. *et al.* In-plane anisotropic and ultra-low-loss polaritons in a natural van der Waals crystal. *Nature* **562**, 557–562 (2018).

27. Zhao, D. *et al.* Recent advances in ultraviolet nanophotonics: from plasmonics and metamaterials to metasurfaces. *Nanophotonics* **10**, 2283–2308 (2021).

28. Novoselov, K. S., Mishchenko, A., Carvalho, A. & Castro Neto, A. H. 2D materials and van der Waals heterostructures. *Science* **353**, aac9439 (2016).

29. Ermolaev, G. A. *et al.* Optical Constants and Structural Properties of Epitaxial $MoS_2$ Monolayers. *Nanomaterials* **11**, 1411 (2021).

30. Yoo, S. & Park, Q.-H. Spectroscopic ellipsometry for low-dimensional materials and heterostructures. *Nanophotonics* **11**, 2811–2825 (2022).

31. Lee, S., Jeong, T., Jung, S. & Yee, K. Refractive Index Dispersion of Hexagonal Boron Nitride in the Visible and Near-Infrared. *physica status solidi (b)* **256**, 1800417 (2019).

32. Passler, N. C. & Paarmann, A. Generalized 4 × 4 matrix formalism for light propagation in anisotropic stratified media: study of surface phonon polaritons in polar dielectric heterostructures. *J. Opt. Soc. Am.*

*B* **34**, 2128–2139 (2017).

33. Stenger, I. *et. al.* Low frequency Raman spectroscopy of few-atomic-layer thick hBN crystals. *2D Mater.* **4** 031003 (2017).

34. Dai, S. *et al.* Tunable Phonon Polaritons in Atomically Thin van der Waals Crystals of Boron Nitride. *Science* **343**, 1125–1129 (2014).

35. Voronin, K., Taradin, A. S., Gorkunov, M. V. & Baranov, D. G. Single-Handedness Chiral Optical Cavities. *ACS Photonics* **9**, 2652–2659 (2022).

36. Taradin, A. & Baranov, D. G. Chiral light in single-handed Fabry-Perot resonators. *J. Phys. Conf. Ser.* **2015**, 012012 (2021).

37. Semnani, B., Flannery, J., Al Maruf, R. & Bajcsy, M. Spin-preserving chiral photonic crystal mirror. *Light Sci Appl* **9**, 23 (2020).

38. Menzel, C., Rockstuhl, C. & Lederer, F. Advanced Jones calculus for the classification of periodic metamaterials. *Physical Review A* **82**, 053811 (2010).

39. Sun, C. *et al.* Single-chip microprocessor that communicates directly using light. *Nature* **528**, 534–538 (2015).

40. Shastri, B. J. *et al.* Photonics for artificial intelligence and neuromorphic computing. *Nature Photonics* **15**, 102–114 (2021).

41. Kresse, G. & Furthmüller, J. Efficient iterative schemes for ab initio total-energy calculations using a plane-wave basis set. *Phys. Rev. B.* **54**, 11169–11186 (1996).

42. Kresse, G. & Furthmüller, J. Efficiency of ab-initio total energy calculations for metals and semiconductors using a plane-wave basis set. *Comput. Mater. Sci.* **6**, 15–50 (1996).

43. Shishkin, M. & Kresse, G. Implementation and performance of the frequency-dependent GW method within the PAW framework. *Phys. Rev. B* **74**, 035101 (2006).

44. Kresse, G. & Joubert, D. From Ultrasoft Pseudopotentials to the Projector Augmented-Wave Method. *Phys. Rev. B.* **59**, 1758 (1999).